\documentclass[sigconf]{acmart}
\renewcommand\footnotetextcopyrightpermission[1]{} 
\def\BibTeX{{\rm B\kern-.05em{\sc i\kern-.025em b}\kern-.08emT\kern-.1667em\lower.7ex\hbox{E}\kern-.125emX}}
    
\copyrightyear{2019}
\acmYear{2019}
\setcopyright{othergov}
\acmConference[ICPP 2019]{48th International Conference on Parallel Processing}{August 5--8, 2019}{Kyoto, Japan}
\acmBooktitle{48th International Conference on Parallel Processing (ICPP 2019), August 5--8, 2019, Kyoto, Japan}
\acmPrice{15.00}
\acmDOI{10.1145/3337821.3337909}
\acmISBN{978-1-4503-6295-5/19/08}

\usepackage{algorithmic}
\usepackage{graphicx}
\usepackage{listings}
\def\BibTeX{{\rm B\kern-.05em{\sc i\kern-.025em b}\kern-.08em
    T\kern-.1667em\lower.7ex\hbox{E}\kern-.125emX}}
    
\newcommand*{\RR}{%
  \textsf{I\kern-.3ex R}%
}

\usepackage[T1]{fontenc}

\begin{document}
%
\title{Holistic Slowdown Driven Scheduling and Resource Management for Malleable Jobs}

%
\author{Marco D'Amico}
\email{marco.damico@bsc.es}
\orcid{0000-0002-6195-6204}
\affiliation{%
  \institution{Barcelona Supercomputing Center}
  \city{Barcelona}
  \state{Spain}
}
\author{Ana Jokanovic}
\email{ana.jokanovic@bsc.es}
\orcid{}
\affiliation{%
  \institution{Barcelona Supercomputing Center}
  \city{Barcelona}
  \state{Spain}
}
\author{Julita Corbalan}
\email{juli@ac.upc.edu}
\orcid{}
\affiliation{%
  \institution{Universitat Politecnica de Catalunya}
  \city{Barcelona}
  \state{Spain}
}
%

%
\begin{abstract}
In job scheduling, the concept of malleability has been explored since many years ago. Research shows that malleability improves system performance, but its utilization in HPC never became widespread. The causes are the difficulty in developing malleable applications, and the lack of support and integration of the different layers of the HPC software stack. However, in the last years, malleability in job scheduling is becoming more critical because of the increasing complexity of hardware and workloads. In this context, using nodes in an exclusive mode is not always the most efficient solution as in traditional HPC jobs, where applications were highly tuned for static allocations, but offering zero flexibility to dynamic executions. This paper proposes a new holistic, dynamic job scheduling policy, Slowdown Driven (SD-Policy), which exploits the malleability of applications as the key technology to reduce the average slowdown and response time of jobs. SD-Policy is based on backfill and node sharing. It applies malleability to running jobs to make room for jobs that will run with a reduced set of resources, only when the estimated slowdown improves over the static approach. We implemented SD-Policy in SLURM and evaluated it in a real production environment, and with a simulator using workloads of up to 198K jobs. Results show better resource utilization with the reduction of makespan, response time, slowdown, and energy consumption, up to respectively 7\%, 50\%, 70\%, and 6\%, for the evaluated workloads.
\end{abstract}

%
%

%

%

%
\maketitle

\section{Introduction and motivation}
In static HPC environments, it is a widely extended practice to allocate full nodes for exclusive utilization. This solution minimizes jobs interferences and applications can be tuned to exploit all the resources available in the node. However, execution environments, architectures, applications and workflows are more and more increasing their complexity. On one side, traditional HPC applications became just a piece of complex puzzles, including, data analytics applications, visualizations tools, creating elaborated workflows. To support those workloads, computing nodes increased their complexity, with the multicore and manycore technology delivering high computing power, in combination with different memory layers for better I/O management. It is difficult for every single piece of the workload to exploit all the available resources, so dynamic resource management approaches need to be evaluated. 

In a context where jobs can share nodes to take advantage of all the available resources, the execution of those workflows could highly improve the hardware efficiency.

Resource sharing as a strategy to improve resource utilization has been considered before with time-sharing approaches such as Gang-scheduling~\cite{gang} or space-sharing, such as oversubscription for co-scheduling~\cite{cosched}, both being not a specific policy but a family of policies in the same way backfill includes many variants. Resource sharing often poorly perform, because of the contention created by multiple applications that, tuned to run in exclusive job allocations, share the same resources, and by the overhead of context switching.

Our proposal overcome this problem by exploiting malleability~\cite{10.1007/BFb0022284} as a way for efficiently managing resources inside computing nodes.
Programming models such as \\OpenMP~\cite{OpenMPspec}, OmpSs~\cite{Duran2011OmpssAP}, and MPI~\cite{MPIspec} give basic support to changing the number of threads or tasks and their binding to resources, but the rest of HPC layers lack communication between them and this feature remain isolated. We started to address this problem by developing DROM, an API~\cite{D'Amico:2018:DEE:3229710.3229752} that enables the scheduler to communicate with applications that can adapt at run time to changes in the computing resources. We took advantage of shared memory programming models to hide programming complexity and to allow dynamic cores allocation efficiently. We also showed that malleability outperform static resource sharing solutions and reduce response time.

In this paper we present SD-Policy, a Slowdown Driven, dynamic job scheduling policy, which exploits malleability offered by DROM as the key technology to reduce the average slowdown and response time of jobs. SD-Policy, based on backfill, applies malleability to running jobs to make room for jobs that will run with a reduced set of resources, only when the estimated slowdown improves over the static scheduling. 
SD-Policy supports mixed workloads with malleable, moldable and static applications, ideal for being used in transition to a malleable environment.
Our approach is holistic, as we put effort connecting all the HPC software stack: applications, programming models, node resource management, and system level scheduling need to work in coordination to achieve the best performance.
We implemented SD-Policy into the SLURM~\cite{10.1007/10968987_3} Workload Manager by extending its plug-ins, and evaluated it with two methodologies: in a real environment, the Marenostrum4 (MN4)~\cite{MN4} system, using up to 49 computing nodes, and in the BSC SLURM simulator~\cite{slurmsim}, using standard workloads~\cite{feitelsonrepo}~\cite{swfweb}, up to 198K jobs on 5040 nodes. With the two methodologies, we demonstrate on one side the feasibility of the SD-Policy in a real production system, and, on the other hand, that significant performance benefits can be obtained with dynamic scheduling policies when executing long, meaningful workloads. Our main contributions are:
\begin{itemize}
\item A dynamic job scheduling, Slowdown-Driven, policy for malleable jobs based on backfill and co-scheduling.
\item An efficient resource selection algorithm based on the feedback of system metrics, i.e., average slowdown.
\item The integration of our proposal in the full software stack: SLURM workflow manager, standard programming models like OpenMP, OmpSs, and MPI, with a negligible effort from developers.
\end{itemize}

The next sections are organized as following: Section~\ref{background} introduces concepts, software, libraries used in our work, Section~\ref{implementation} describes the developed job scheduling policy and its implementation in a well know workload manager. Section~\ref{evaluation} presents both simulations and real runs evaluations on the presented policy and its parameters, together with an in-depth analysis of their effects on system performance. Finally, Section~\ref{RW} present the state of the art of the subject of scheduling in the context of malleable jobs, and Section~\ref{future} resumes and concludes our research, with insight on future work.

\section{Background}\label{background}

\subsection{Malleability and DROM}
A malleable job is able to adapt to changes in the allocated resources dynamically at runtime\cite{feitelsonrepo}. For modern HPC hardware architectures, it is important to divide malleability into two parts: malleability among different resources available in the nodes, and malleability in the number of computing nodes. 

In previous work, we developed DROM (Dynamic Resource Ownership Management)~\cite{D'Amico:2018:DEE:3229710.3229752}, an interface that puts in communication applications and node managers, enabling the first level of malleability. DROM avoids the high overhead of re-configuring and transferring data needed by the second approach, as explained in the related work, in Section~\ref{RW}. 

The interface enables malleability for jobs by integrating with shared memory programming models, like OpenMP~\cite{OpenMPspec} and \\OmpSs~\cite{Duran2011OmpssAP}. It is designed to be integrated with any programming model, or directly with applications, enabling shrink and expand operations in malleability points, e.g., between tasks execution, where it is safe to change used resources. DROM eliminates overhead for developers that only need to dynamically link the library, without modifying or recompiling the code.
The library also implements API for registering processes in the DROM environment, getting the list of recorded processes, and getting/setting their CPU masks.
Those API can be used by the resource manager, in particular by the node manager, that is in charge to communicate with applications when the scheduler asks to change the job's allocation.

We evaluated the overhead of DROM in different use cases, and we found that the cost of shrinking and expanding operations are negligible, while response time and resource utilization improve. The overhead is given by higher level cache invalidation and additional code injected by the library every time a job shrinks or expands.
Motivated by positive results, we decided to investigate the benefits of widely using DROM in an HPC system, e.g., by malleable-backfilling relatively short jobs, that complete fast, to improve the system slowdown. Low overhead allows SD-Policy to apply malleability at high frequency, avoiding at the same time to penalize only some specific jobs. Using the second level of malleability at this granularity would require more significant amounts of time just to perform the shrinking and expanding operations, voiding the potential benefits on the system.
On the other hand, our policy is compatible with malleability in the number of computing nodes, that applied at a different granularity, could bring of further improvements in system performance.

\subsection{SLURM and SLURM simulator}\label{slurmsim}
SLURM workflow manager~\cite{10.1007/10968987_3} is an open source, efficient job manager, popular in research and widely used in HPC systems, installed on 6 of the top 10 supercomputers of the top500 list.

SLURM is based on a master-slave architecture, with a master controller called \textit{slurmctld} and a slave daemon for each compute node, called \textit{slurmd}.
Slurmctld is the central coordinator, it receives requests from users that submit jobs, it manages jobs in a priority queue, and it includes scheduling and placement algorithms for launching jobs in compute nodes. 
Slurmd is the node manager, it communicates with slurmctld to receive job launch requests, and it manages job's launch and supervises their execution.

BSC SLURM simulator~\cite{bscsimulator} is based on the SLURM code, plus additional modifications that permit to simulate jobs instead of running them. In a previous work~\cite{slurmsim} we extended and improved the simulator to a stable and more accurate version.

SLURM simulator allowed us to reuse the implemented code into real SLURM, and at the same time, it gave us the possibility to keep using SLURM parameters to have a more precise evaluation, similar to real runs experiments. We developed and integrated into the simulator a runtime model for malleable jobs, as we describe in Section~\ref{simimplementation}.

\section{The Slowdown Driven policy}\label{implementation}
This section presents Slowdown Driven policy and its implementation~\cite{sdpolicycode} in SLURM Workload Manager.

SD-Policy is a variant of backfill, co-scheduling malleable jobs in non exclusively allocated nodes, where they can efficiently partition the available resources using malleability.
It is based on the simple idea that if an arriving job is malleable and not enough resources are available to run it by static backfill algorithm, SD-Policy selects some of the already running jobs, called mates, shrinking them to run the new job, only if predicted system slowdown is improved. Mates, selected by minimizing their increase in slowdown, will be expanded back when the new scheduled job terminates.


We divide the SD-Policy in three parts: \textit{scheduling level}, \textit{resource selection level}, and \textit{node level}.
We will start describing the scheduling policy in the controller, from a system level point of view. Then we will present the implementation of the resource selection and placement algorithm, based on the impact on job mates, and feedback from system metrics, used by the scheduler. Finally, we will explain how malleable applications interact with DROM and the node manager, and the implemented logic for shrinking and expanding operations, cores selection and distribution. At the end of the chapter, we present the used runtime models for the scheduling algorithm and the simulator.


\subsection{The scheduling algorithm}\label{sched_policy}
The scheduling algorithm is a variant of backfill, that considers malleability. It first tries static placement of each job, and if not enough free resources are available, it attempts the flexible scheduling approach for the same job. Malleable backfill runs for each job right after the static trial, and not after the static backfill completed for all jobs. This strategy favors the scheduling of jobs in order of priority. The scheduling algorithm is detailed in Listing~\ref{alg:sched}.

\begin{lstlisting}[basicstyle=\small,language=Python, caption={SD-Policy scheduling algorithm}, label={alg:sched}]
schedule(new_job)
    j = new_job
    if(!(nodes = select_nodes(j,free_nodes,null)) 
        if(!malleable(j))
            return
    else run_job(j, nodes)
    static_end = get_wait_time(j) + j.req_time
    mall_end = j.req_time + runtime_increase(j)
    if (static_end > mall_end)
        s_mates = select_nodes(j,free_nodes,nodes)
        if(s_mates)
            update_stats(j,s_mates)
            s_nodes = get_nodelist(s_mates)
            run_job(j,s_nodes)
\end{lstlisting}

The scheduler uses an end time prediction, \textit{static\_end} and \\\textit{mall\_end}, to estimate if applying malleability would improve the new job slowdown, and, in the affirmative case, it calls the malleable resource selection algorithm. End times are calculated using an estimation of the wait time in the static case, by creating a map of jobs reservations in time to find out when the new job will start. In the malleable case, the job will begin immediately, while runtime will be calculated by adding Equation~\ref{eq:worstcasemodel} to the requested time, from runtime models described in Section~\ref{simimplementation}.


The scheduler collects the list of selected mates by calling \textit{select\_nodes}, using the algorithm described in Section~\ref{rm_policy}, and for each mate and the new job, it updates the requested time and the predicted end time. Then it communicates with the node managers, starting procedure described in Section~\ref{node_policy}.

We modified SLURM backfill scheduler, by editing \textit{sched/backfill} plug-in, to take advantage of malleability by implementing Listing~\ref{alg:sched}.


\subsection{The resource selection algorithm}\label{rm_policy}
The resource selection problem, when co-scheduling multiple jobs with malleability, is reduced to selecting best job mates that will shrink their allocation to make room for the new jobs. 
Selecting mates is a NP-complete problem, the objective function tries to find the best set of mates with minimum penalty \textit{p}, i.e. the jobs that receive minimum performance impact when malleability is applied. We describe it with Equations~\ref{eq:objective}-\ref{eq:constraint2}.
Given:
\begin{itemize}
\item $x_i \in \{0,1\}$: mate \textit{i} is selected
\item $n$ number of mates
\item $p_i$ penalty estimated for mate
\item $P$ maximum penalty for a single mate
\item $w_i$ weight of the mate \textit{i}
\item $W$ weight of the scheduled job
\end{itemize}
we define the \textit{Performance Impact (PI)} with the objective function~\ref{eq:objective}:

\begin{equation}\label{eq:objective}
PI = min \sum\limits_{i=1}^n x_i * p_i
\end{equation}

With the constraints~\ref{eq:constraint1} and~\ref{eq:constraint2}:

\begin{equation}\label{eq:constraint1}
p_i < P, \forall i \in \RR
\end{equation}

\begin{equation}\label{eq:constraint2}
\sum\limits_{i=1}^n x_i * w_i = W
\end{equation}

Following on, we define the Performance Impact \textit{PI}, penalties \textit{p} and \textit{P}, weights \textit{w} and \textit{W}, and the heuristic used to solves this problem.

\subsubsection{Performance Impact}
\textit{PI} is defined as the sum of slowdown increase for each mate when malleability is applied. As Feitelson~\cite{10.1007/BFb0053978} reports, there is no best metric when we talk about job scheduling evaluation, but it seems slowdown metric helps faster convergence in the preemptive scheduling, similar to our policy. Slowdown, normalizing response time by the runtime, does not give precedence to long jobs like response time, increasing fairness for users submitting short to medium jobs.

\subsubsection{Penalties}
We calculate the penalty \textit{p} assigned to each mate based on estimation of the slowdown increase as:
\begin{equation}\label{eq:penalty_sd}
p_i  = (wait\_time + increase + req\_time) / req\_time
\end{equation}

Equation~\ref{eq:penalty_sd} considers \textit{wait time}, increase in total run time based on Equation~\ref{eq:worstcasemodel}, \textit{increase}, and the user \textit{requested time} for the job. Equation~\ref{eq:penalty_sd} is an estimation, as the job duration is usually not equal to the requested time.

The penalty will give precedence to jobs that waited less in the queue and jobs that request a larger amount of time, so the impact in slowdown will be minimum. We define \textit{P} as \textit{MAX\_SLOWDOWN}, a cut-off for \textit{p}. A cut-off is needed for two main reasons: reducing the eligible mates to reduce the computation, and avoid penalizing jobs that have a high slowdown. We implemented this parameter in two ways:
\begin{enumerate}
\item \textit{A static value} chosen by system administrators. The value can be chosen empirically or by analyzing the history of a system. In this case, the slowdown must be calculated by using user time estimation, not the real runtime, because this metric is the only one the system can use to predict the slowdown of running and waiting jobs.
\item \textit{A dynamic value}: the cut-off is automatically set by the scheduler based on system average slowdown of running jobs, and it is updated every time the controller is not busy in scheduling jobs. The basic idea is to spread the slowdown in a similar way among running jobs, so jobs exceeding the average slowdown are not considered for malleability. Other metrics, like median and 70 percentile were analyzed, but they did not report improvement overall.
\end{enumerate}

\subsubsection{Weights}
Following constraint~\ref{eq:constraint2}, we define weight \textit{w} as the number of allocated computing nodes for the mate, and \textit{W} as the number of computing nodes requested by the scheduled job. Constraint~\ref{eq:constraint2} helps to keep balanced the system in the number of cores per node jobs use in
the allocated nodes, assuming jobs are statically load balanced, which is the common case in HPC environment.

\subsubsection{Heuristic}
The proposed heuristic tries different combinations of mates iterating recursively on the list of mates for the power of \textit{m} times, where \textit{m} is a configurable parameter representing the maximum number of mates. From our evaluation with standard workloads we did not see improvements in system metrics increasing \textit{m} over two, so we kept it as an optimal value. For each combination that satisfies constraints~\ref{eq:constraint1}-\ref{eq:constraint2}, we calculate the Performance Impact and update the best solution, if improving it.
Mates list is sort based on the mates' penalty \textit{p}. If the list is too big, it can be reduced by considering only the first \textit{nm} mates, with lower \textit{p}. The algorithm supports contiguous allocations, node filtering by name, architecture, memory and network constraints. Options such as including free nodes to reduce fragmentation and more than two mates per node are supported.

We set a further constraint: new jobs must finish inside mates' allocation to avoid, in the case of multiple mates, that one of them finishes earlier. We avoid this case because the new job would expand to occupy the full nodes of just a part of its allocated space, creating unbalance in the case the application cannot balance its load dynamically.
The constraint also avoids creating a further delay for jobs scheduled to run afterward.  
Listing~\ref{alg:nodeselect} synthesizes the presented algorithm.

\begin{lstlisting}[basicstyle=\small,language=Python, caption={SD-Policy node selection heuristic}, label={alg:nodeselect}]
select_nodes(new_job, nodes, free_nodes)
    if (free_nodes.count > new_job.requested_nodes)
        return s_nodes = pick_nodes(new_job,free_nodes)
    if(nodes && malleable(newjob))
        mates = get_list_of_mates()
        mates = filter_and_sort(mates,MAX_SLOWDOWN)
        s_mates = pick_mates(newjob,mates,nodes)
        return s_mates
\end{lstlisting}

We implemented the malleable resource selection algorithm in SLURM \textit{select/linear} plug-in. This plug-in is in charge to select entire nodes for jobs to be scheduled, respecting all job's constraints and optimizing the placement of the nodes using different criteria.
We opted for the linear plug-in because malleable jobs can expand and shrink in the node, so there is no need to select individual cores at this point of the scheduling flow. The plug-in allocates entire nodes to give the node manager more freedom in binding specific cores to jobs, as it has a better view of the node usage, and it can get information directly from applications.

\subsection{Node management algorithm}\label{node_policy}
Node management is the bottom layer of resource management, and it directly interacts with jobs. In the SD-Policy, node managers are able to select the right amount of computing cores to give to each job, assigning them at launch time, and also controlling their number at runtime.

To achieve described malleability we integrated DROM API into the node manager, to automate the placement of jobs' tasks inside computing nodes whenever one or more malleable jobs are co-scheduled in overlapping job's allocations. 
Node managers calculate tasks to cores distribution among jobs and automatically, keeping jobs balanced and isolated. Exceptional cases with nodes running different configurations or jobs asking a different distribution are supported, e.g., using a manual organization to optimize usage of resources for master-slave application architectures, or memory intensive jobs that do not need a high number of computing cores, but instead more memory bandwidth. In the second case, particularly common in the described HPC workloads, a distribution with few computing cores per socket will leave more resources to mates jobs, while fully taking advantage of the memory bandwidth.

We defined the \textit{SharingFactor}, a limit on computational resources that can be taken from a running job in a computational node when shrunk, to implement fairness for mates and to study performance when changing the number of assigned resources.  We evaluated static values for this parameter in Marenostum IV~\cite{MN4} and different cores distributions algorithms for the automatic distribution case. Results show that best overall performance is obtained when the applications run isolated in separate sockets. For Marenostrum, the number of sockets is two, so the sharing factor is set to 0.5.
In a dynamic approach, online performance analysis of running jobs would feed a tuning algorithm for selecting optimal values of SharingFactor, further increasing nodes efficiency.

The number of cores assigned to the malleable scheduled job depends on the SharingFactor and the minimum amount of cores to which the running jobs can shrink. This last value is equal to the static number of MPI ranks the application is running, to which we assign a minimum of one computing resource per rank.

\begin{lstlisting}[basicstyle=\small,language=Python, caption={SD-Policy node management algorithm}, label={alg:nodemngr}]
while(1)
    if(new_job = get_next_job())
        if(malleable(newjob))
            runningjobs.add(new_job)
            distribute_cpu(runningjobs)
            for job in running_jobs
                shrink_job(job)
            DROM_run(newjob)
        run(newjob)
    if(end_job = get_finished_job())
        if(malleable(end_job))
            if(end_job == mate)
                distribute_cpu(runningjobs)
            else
                owner = get_owner(end_job, runningjobs)
                expand_job(owner)
            DROM_clean(end_job)
\end{lstlisting}

The implementation is enclosed in the SLURM's \textit{task/affinity} plug-in, in charge of controlling the resources assigned by \textit{slurmctld} to the job's tasks. Task/affinity is dynamically loaded by \textit{slurmd} and \textit{slurmstepd}, dividing the code flow into two parts. In the first part, \textit{Slurmd}, the node manager, is in charge of managing computing resources of a specific node, and thanks to the plug-in, calculating the affinity of tasks, i.e., the distribution onto the computing cores. \textit{Slurmstepd}, the job step manager, in the second part, controls the correct task's launch and execution.

The description of the algorithm presented in Listing~\ref{alg:nodemngr}, and its implementation in SLURM follows:
\begin{enumerate}
\item At job start or end, the node manager (slurmd) interacts with DROM to get information about running tasks. In the case of a starting job, our algorithm recalculates affinities for all the tasks in the node. Cores distribution intelligently maintains running and new processes balanced in the number of cores per task, assuming that, without any other information, the imbalance degrades performance. Cores distribution keeps jobs in separate sockets to improve data locality and reduce interference between jobs. At job's end, the algorithm returns cores of the ended job to the owner. If any of the already running owner terminates its execution before the new job, its cores will be distributed to remaining running tasks, to increase node utilization.
\item Once calculated cores distribution, the job manager (slurmstepd) checks if the dependencies created by redistribution of cores are satisfied. In other words, it checks if the jobs where the new job will take cores from are already running or about to start, to assure that DROM assigns cores correctly. Afterward, the job manager launches tasks using the DROM API.
\item At job's launch time, DROM launches and sets the affinity for each new task and attaches them to the DROM space. At end time, it cleans information of tasks from the DROM space. DROM also sets the new affinities for running tasks, so when the tasks reach the next malleability point they can adapt to the change. At end time the API can be set to return cores to the original owners.
\end{enumerate}

\subsection{Runtime model for malleable jobs}\label{simimplementation}
SD-Policy is based on time estimations, so we implemented a model shaping how applications' duration is affected when malleability is applied. We already showed that performance impact of DROM applications when changing the number of resources is negligible, so we assume the duration of jobs is proportional to the number of assigned computing resource, and the ability of jobs to adapt to load unbalance.

To estimate the increase in the runtime of a shrunk job, we partitioned the job duration while sharing nodes in \textit{T} time slots \textit{t}, each time slot is a different job's resource configuration. We proportioned the runtime to the number of used resource for each \textit{t} by putting in relation the static duration with the number of resources.
We developed two models: the \textit{ideal} and the \textit{worst case}.

In the \textit{ideal model}, applications do not suffer from the imbalance in the number of resources used by each process. If one job's task uses one computing node and the other task just half, with this model, the performance will be linear with the number of resources. In this case we compute the increase in the runtime with Equation~\ref{eq:idealmodel}.

\begin{equation}\label{eq:idealmodel}
increase = \sum\limits_{t=1}^T (req\_cpus / used\_cpus_t * time_t)
\end{equation}
This model represent applications that can dynamically adapt their load to the resources at run time. \textit{tot\_cpus} is the original number of cores used in the log, \textit{used\_cpus} the assigned number at each time slot.

In the \textit{worst case model} the same scenario would bring to lower utilization since performance is limited by the less used node \textit{n} over all nodes \textit{N}. We compute the increase in the runtime with Equation~\ref{eq:worstcasemodel}.
\begin{equation}\label{eq:worstcasemodel}
\begin{aligned}
increase = \sum\limits_{t=1}^T (req\_cpus / \min{(cpus\_per\_node_{nt})} * time_t)\\
\forall n \in \{1, N\}
\end{aligned}
\end{equation}
The second model represents statically balanced applications. In this case, unbalanced changes in the used resources among nodes generate imbalance.
Characterization of applications would led to a more detailed model that would improve simulated experiments precision, but in its absence considering the two models, we are able to give an upper and a lower bound.

We used the described models for time estimations in the SD-Policy and to calculate jobs' runtime in the SLURM simulator. In the SD-Policy case, we use the \textit{worst case model}, to be able to grant correct jobs execution and completion.
In the simulator we try both models, and compare results, in Section~\ref{ideal_worst}. Running real runs experiments, in Section~\ref{realrun} will be the third source of information on how typical HPC applications perform when adapted to be malleable.


\section{Evaluation}\label{evaluation}
Our evaluation is divided into two parts, depending on the used methodology: workloads run in a real environment and workloads simulations. 
We use the first approach to give a proof of the effective implementation of the policy running in a production system, and its integration with well-known schedulers and programming models, as well as a performance evaluation with benchmarks, neural network spiking simulators coming from Human Brain Project~\cite{HBP}, and a computational multi-physics solver. We evaluate this workload on Marenostrum4 (MN4)~\cite{MN4} supercomputer, on up of 49 computing nodes with two sockets equipped with Intel Xeon Platinum 8160 processors, 48 cores, 96GB of main memory per node, for a total 2353 cores. Each workload runs for about two days.

The second methodology, based on whole systems simulations, allowed scaling the evaluation to workloads up to 80640 cores for eight months, allowing the analysis of the SD-Policy performance running on entire HPC systems.

Table~\ref{workloads_desc} presents information about workloads used for the evaluation. All workloads and models come from Feitelson database web page~\cite{feitelsonrepo}. We generated workloads 1, 2 and 5 with the model developed by Cirne~\cite{cirne}, based on the characterization of four different logs. We configured it to use ANL arrival pattern, and we scaled the model to the considered system size.
Since we were interested in the performance of our algorithm when the time-based predictions are precise, we generated \textit{Workload 2}, equal in distribution to \textit{workload 1} but with the job's requested time same to the real duration.
\textit{Workload 3} is part of the RICC installation trace from 2010, a realistic workload characterized by a high number of small jobs requesting few nodes, ranging from short to long runtime, up to four days.
\textit{Workload 4} is the cleaned version of CEA-Curie log from 2011, only considering the primary partition.

\textit{Workload 5} was created from Cirne model, then converted to real applications submissions. We modeled applications behavior and scalability presented in Table~\ref{table:apps_real_run}, then we calculated parameters to adapt the applications to each entry of the workload, regarding the number of requested nodes and duration of the job. We chose a set of applications with different behaviors in the utilization of CPUs and main memory. In the lack of statistics about HPC workloads characterization in the literature, we organized the workload in three main types of applications, equally distributed: compute-bounded jobs with low memory utilization (PILS), memory-bounded jobs with lower CPU utilization (STREAM), big simulations, memory and compute intensive (CoreNeuron, NEST, Alya).

\begin{table*}[htp]
\centering
\caption{Description of workloads}
\label{workloads_desc}
\begin{tabular}{|c|c|c|c|c|c|c|c|}
\hline
\textbf{ID} & \textbf{Log/model} & \textbf{\# jobs} & \textbf{System (nodes/cores)} & \textbf{max job (nodes/cores)} & \textbf{Avg resp. time (s)} & \textbf{Avg slowdown} & \textbf{Makespan (s)} \\ \hline
1 & Cirne              & 5000             & 1024/49152                       & 128/6144                          & 122152                         & 3339,5                & 899888                          \\ \hline
2 & Cirne\_ideal       & 5000             & 1024/49152                       & 128/6144                          & 126486                         & 3501                  & 896024                          \\ \hline
3 & RICC-sept     & 10000            & 1024/8192                        & 72/576                            & 43537                          & 1341                  & 407043                          \\ \hline
4 & CEA-Curie          & 198509           & 5040/80640                       & 4988/79808                        & 29858,5                        & 3666,5                & 21615111                        \\ \hline
5 & Cirne\_real\_run & 2000             & 49/2352                          & 16/768                            & 56482                           & 4783,1                  & 159313                            \\ \hline
\end{tabular}

\end{table*}

\begin{table*}[htp]
\centering
\caption{Workload characterization for real runs evaluation}
\label{table:apps_real_run}
\begin{tabular}{|c|c|c|c|c|c|}
\hline
Application & \% workload & ReqNodes      & ReqTime       & CPU utilization & Memory utilization \\ \hline
PILS~\cite{hints}        & 30.5\%           & small to high & small/med     & high            & low                \\ \hline
STREAM~\cite{stream}      & 30.8\%           & small to high & small/med     & low             & high               \\ \hline
CoreNeuron~\cite{coreneuron}  & 35,5\%           & small to high & small to high & high            & med                \\ \hline
NEST~\cite{nest}        & 2.6\%            & small to high & small to high & high            & med                \\ \hline
Alya~\cite{VAZQUEZ201615}        & 0.6\%            & small         & high          & high            & med                \\ \hline
\end{tabular}%

\end{table*}

The following section presents different evaluations:
\begin{itemize}
\item Evaluation of MAX\_SLOWDOWN
\item Analysis of simulated workload 4
\item Evaluation of runtime models
\item Analysis of Workload 5 in a real environment
\end{itemize}

We used the following metrics for the evaluation:
\begin{itemize}
\item Makespan: the difference between the last job end time and the first job arrival time.
\item Average response time: the average of jobs' response time, calculated as the difference between jobs' end time and submission time.
\item Average slowdown: the average of jobs' slowdown, calculated as the response time divided by the static execution time of the job.
\item Energy consumption: The energy consumed to run entire workloads, reported by system software.
\end{itemize}

\subsection{Evaluation of MAX\_SLOWDOWN}
Different values of MAX\_SLOWDOWN can have a high impact on the performance of SD-Policy. Low values will limit the number of times malleability can be applied, and high values could degrade jobs and system performance.
We simulated workload 1, 2, 3 and 4 using SharingFactor of 0.5 and the ideal runtime model. We tried different values for MAX\_SLOWDOWN represented by the following labels: MAXSD 5, MAXSD 10, MAXSD 50, MAXSD infinite, and the dynamic cut-off based on feedback from running jobs slowdown, DynAVGSD.

\begin{figure}[htbp]
  \centering
  \includegraphics[width=0.9\columnwidth]{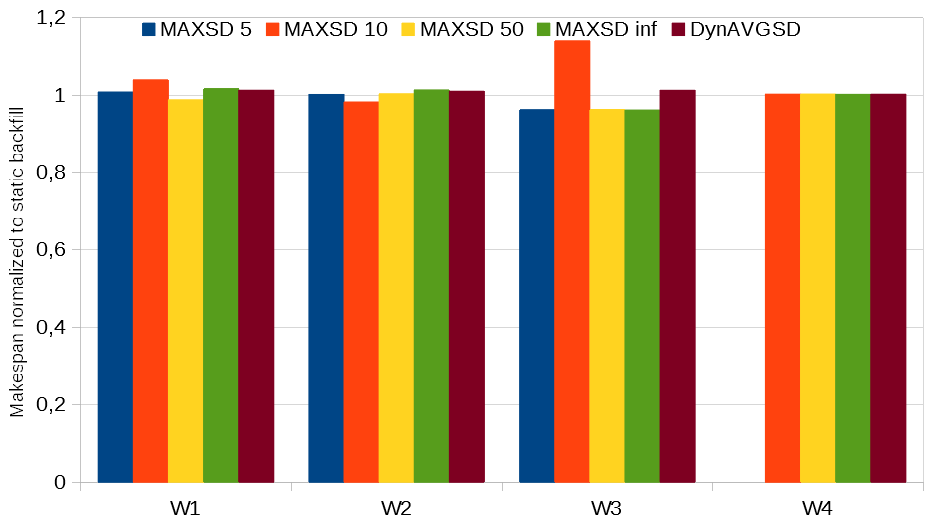}
  \caption{Makespan for workload 1 to 4 changing the MAX\_SLOWDOWN parameter, normalized to static backfill simulation.}
  \label{fig:wallclock_maxsd}
\vspace{0.2cm}
  \centering
  \includegraphics[width=0.9\columnwidth]{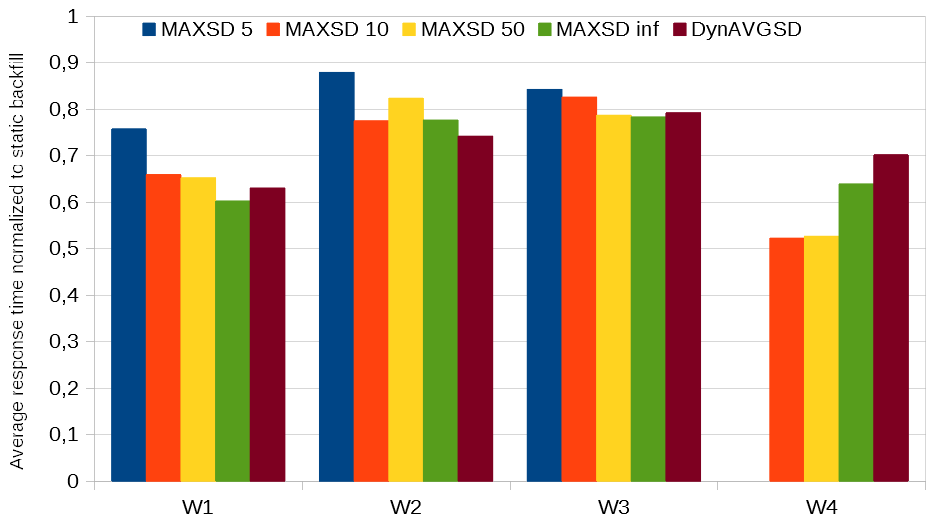}
  \caption{Average response time for workload 1 to 4 changing the MAX\_SLOWDOWN parameter, normalized to static backfill simulation.}
  \label{fig:resptime_maxsd}
  \vspace{0.2cm}
  \centering
  \includegraphics[width=0.9\columnwidth]{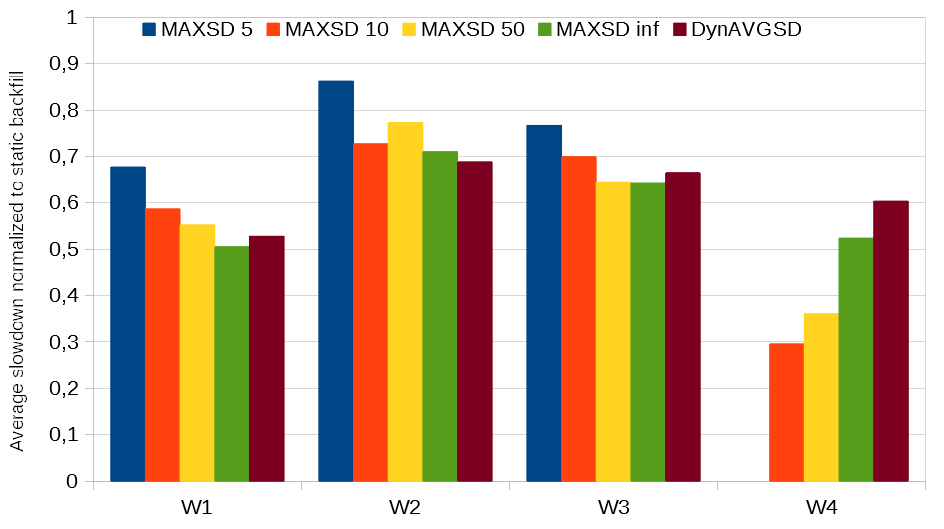}
  \caption{Average slowdown for workload 1 to 4 changing the MAX\_SLOWDOWN parameter, normalized to static backfill simulation.}
  \label{fig:sd_maxsd}
  \vspace{-0.5cm}

\end{figure}
Figures~\ref{fig:wallclock_maxsd},~\ref{fig:resptime_maxsd}, and~\ref{fig:sd_maxsd} show the makespan, average response time and average slowdown for the workloads, normalized to static backfill.
In the first three workloads, increasing slowdown limit improves the system's slowdown, showing that not filtering mates still improves the system performance overall. We observe a reduction in slowdown up to 49.5\%, 31\%, 25.7\%, and 70.4\% respectively for Workload 1, 2, 3, and 4.

On one side, even with a high limit for MAX\_SLOWDOWN, our algorithm tries to avoid applying malleability when it would not bring improvements, avoiding performance loss also when increasing the parameter to an infinite value. On the other side, a system administrator could still consider using a relatively low limit, to avoid some jobs to be penalized too much, or implement different queues with different QoS policies using different MAXSD configurations.

In Workload 2 slowdown does not monotonically decrease with the increase of the limit, showing an increase at value 50, but still outperforming static backfill. SD-Policy DynAVGSD brings further improvements in the same workload, where the SD-Policy works with real jobs durations and not the user requested time. The explanation of the observed behaviors resides in the fact that variance in the real average slowdown is much higher than in the predicted average slowdown when using user requested time, so a dynamic value of MAX\_SLOWDOWN benefits this evaluation. Also, using real predicted metrics allows the SD-Policy to be more precise. This observation suggests that using a predictive method for job's runtime, i.e. based on machine learning, rather than asking the user, will improve the performance of our policy.
Comparing workload 1 and 2, the static backfill for the first workload outperforms the static backfill for the second by 8.6\%, showing the interesting result that precision of job's duration does not always produce better average system slowdown. On the other side, having an exact job duration allows jobs not to be delayed in their start time, making backfill correct. Backfill behavior influences our approach, also base algorithm of SD-Policy, with Workload 1 performing 29.6\% better than Workload 2.

In Workload 4, the best value is obtained in the MAXSD 10 case, while higher values and the dynamic version do not improve results. For this significant workload, we have the maximum observed reduction of the response time and the slowdown up to respectively 50\% and 70\%, showing the high benefits a system can have over a static backfill approach.

\subsection{Analysis of a big workload}
\begin{figure}[htbp]
 \centering
 \includegraphics[width=1\columnwidth]{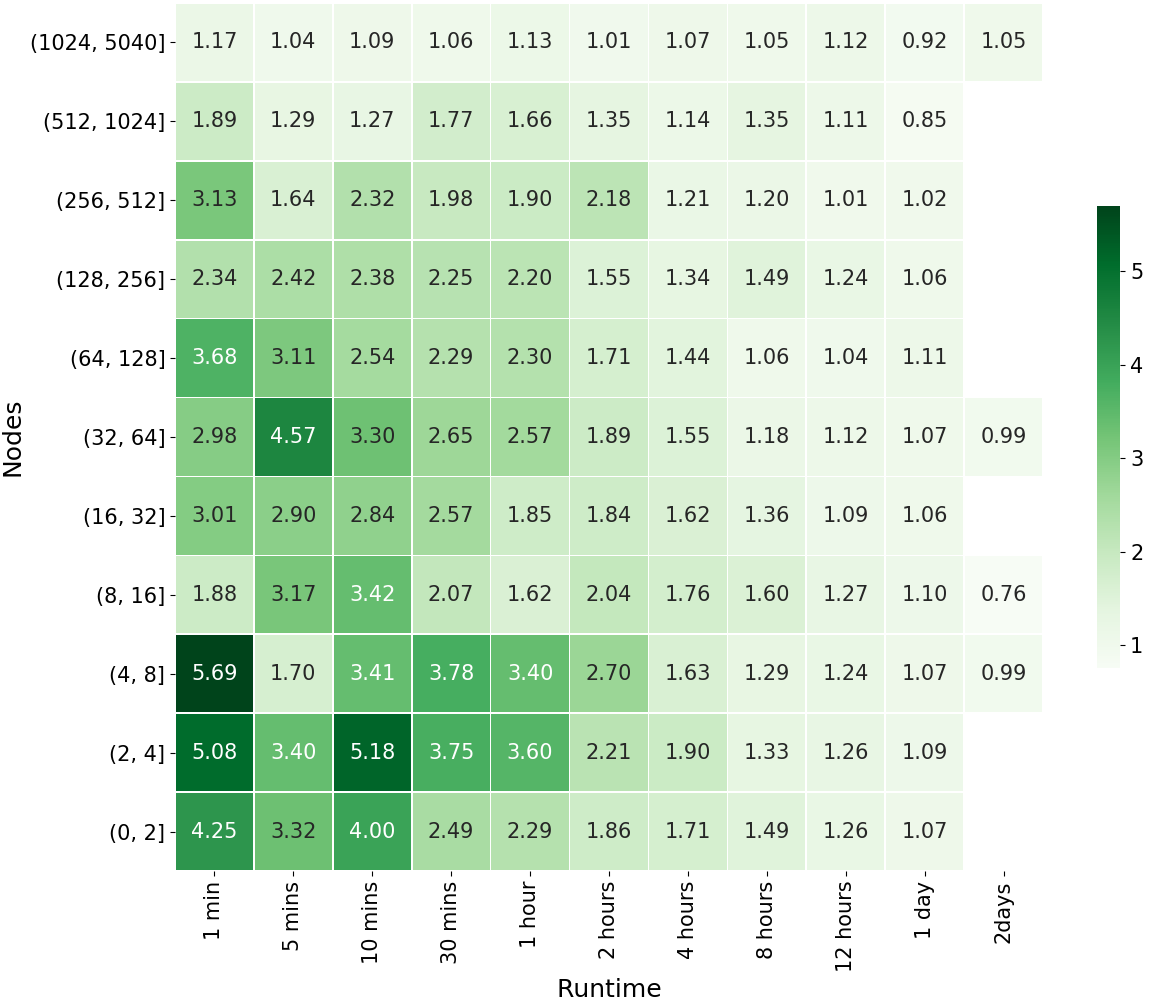}
 \caption{Heatmap showing the ratio between slowdown of static backfill and SD-Policy MAXSD 10 using workload 4, for different job categories.}
 \label{fig:cea_sd_hm}
 \vspace{0.2cm}
 \centering
 \includegraphics[width=1\columnwidth]{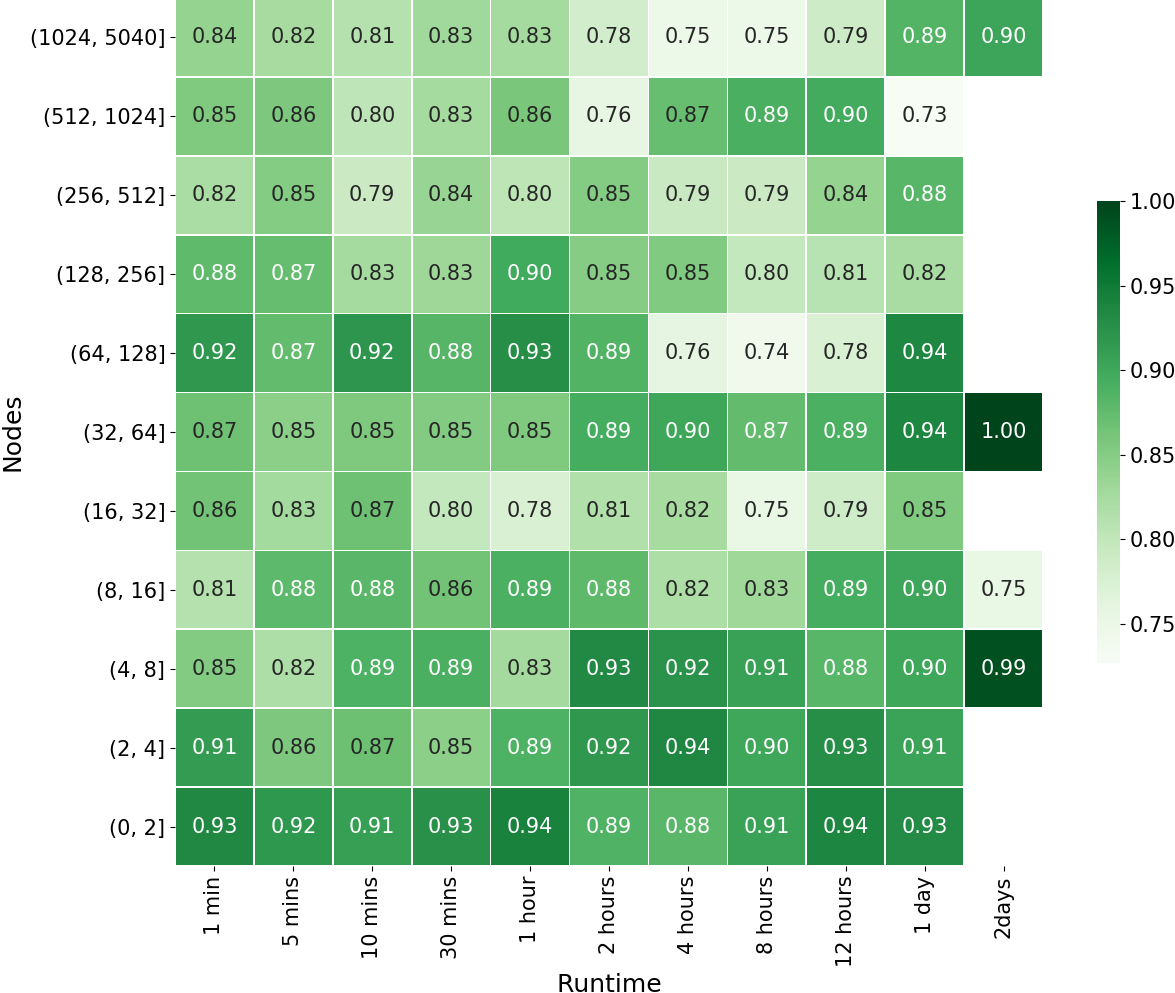}
 \caption{Heatmap showing the ratio between run time of static backfill and SD-Policy MAXSD 10 using workload 4, for different job categories.}
 \label{fig:cea_rt_hm}
 \vspace{-0.3cm}
\end{figure}

\begin{figure}[htbp]
 \centering
 \includegraphics[width=1\columnwidth]{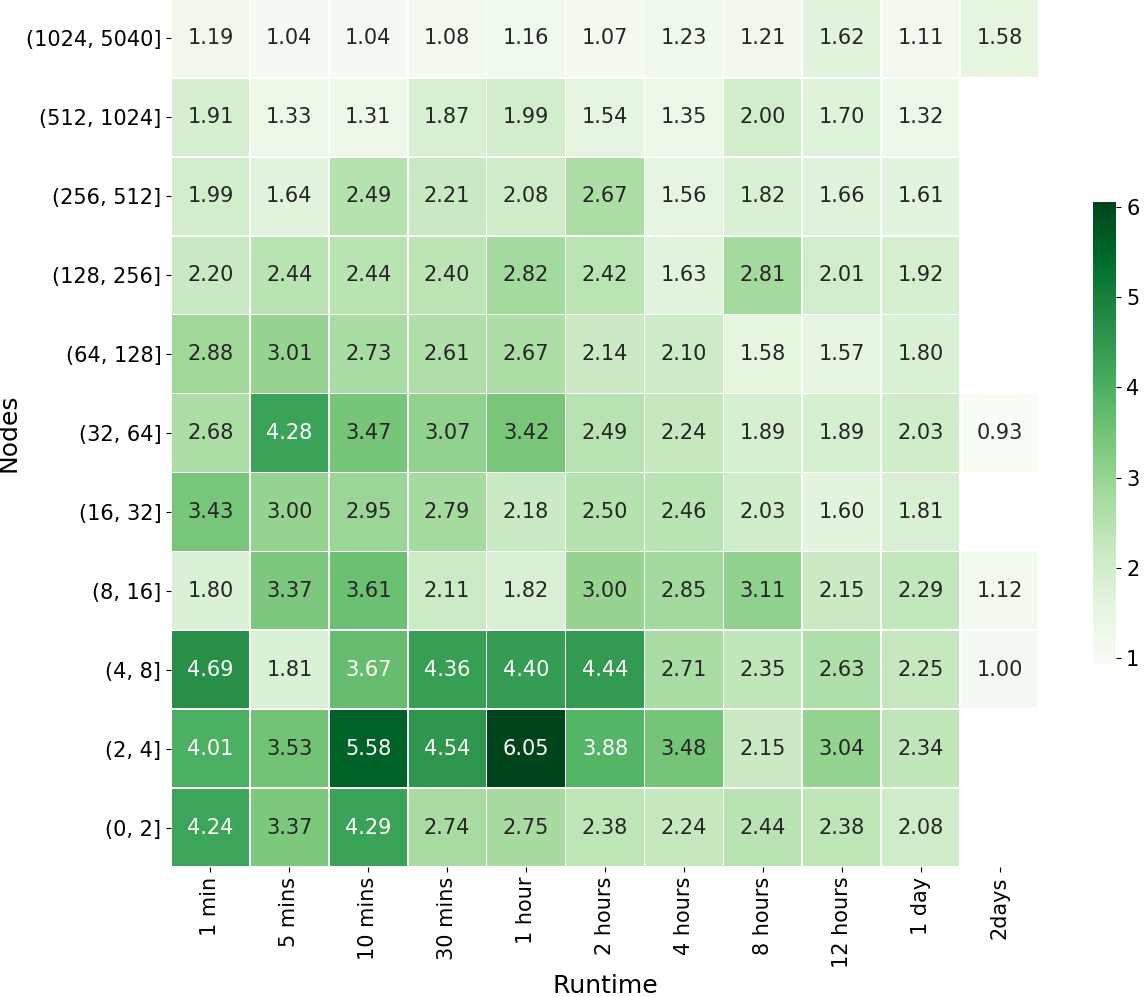}
 \caption{Heatmap showing the ratio between wait time of static backfill and SD-Policy MAXSD 10 using workload 4, for different job categories.}
 \label{fig:cea_wt_hm}
\vspace{0.2cm}
 \centering
  \includegraphics[width=1\columnwidth]{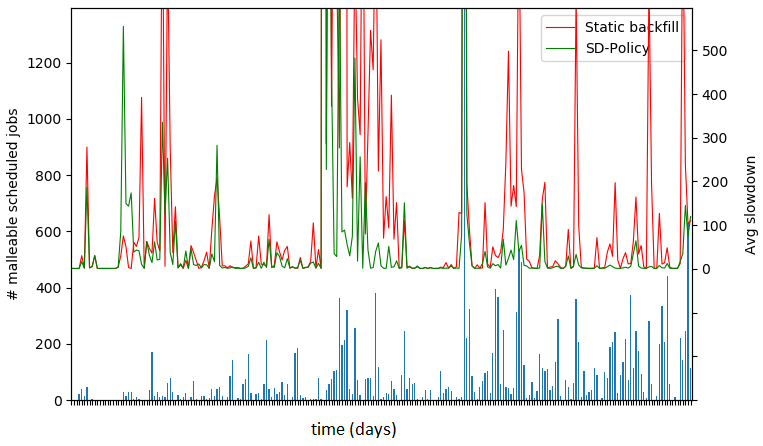}
 \caption{Columns represent the number of jobs scheduled with malleability per day, their y axis is on the left. The two lines represent slowdown per day of static backfill and SD-Policy MAXSD 10.}
 \label{fig:cea_nmall_avgsd}
 \vspace{-0.3cm}
\end{figure}

We simulated workload 4, big in length and number of nodes, and analyzed details by partitioning the jobs in categories depending on the requested resources and runtime. We compared the static backfill version with the MAXSD 10 version. The SD-Policy improved slowdown by 70.4\% while keeping makespan constant. We compared average slowdown, average runtime and average wait time for static backfill and the SD-Policy, presenting the ratios between the two versions in Figures~\ref{fig:cea_sd_hm},~\ref{fig:cea_rt_hm} and~\ref{fig:cea_wt_hm}. The malleable version highly improves the slowdown of jobs consuming up to 4 hours and asking up to 512 nodes, up to 569\%. Relatively small and short jobs have very high slowdown, compared to larger and longer jobs. Those are the primary jobs that benefit from the SD-Policy.

The rest of the heatmap of the slowdown, in Figure~\ref{fig:cea_sd_hm} keeps having better values even for bigger and larger jobs, except for three categories. Two categories contain few jobs to take some conclusions, but the 121 jobs asking 512 to 1024 nodes with duration in 12 hours to 1 day range, show an increase of 15\% in the average slowdown. Even if it seems this job category is penalized, the average slowdown for those jobs was lower than other neighbors in the heatmap, so result shows that the SD-Policy generates a more fair distribution of the slowdown with respect to the static backfill. The rest of the jobs, even if affected in their runtime (Figure~\ref{fig:cea_rt_hm}), because of the malleability, improve their wait time (Figure~\ref{fig:cea_wt_hm}), showing fairness is not an issue for some particular category of jobs.

Figure ~\ref{fig:cea_nmall_avgsd} shows the trend of the average slowdown per day, comparing the static backfill and the SD-Policy per day, together with the number of jobs scheduled with malleability. It is visible that the peaks in slowdown are highly reduced all over the simulated time, and, apart from an initial spike of jobs that are selected as mates, the average slowdown for our policy never increases over the static. reduction of peaks in slowdown are usually associated with an high number of jobs scheduled with malleability. The total number of malleable scheduled jobs is 20476, the number of mates is 17102, corresponding to 10.3\% and 8.6\% of the workload.

\subsection{Evaluation of runtime models}\label{ideal_worst}
When one of the mates completes before the user requested time, it could lead the other job running in the same node to take the freed cores, but only for part of its job allocation, creating unbalance in its load. We run simulations with the two runtime models presented in Section~\ref{simimplementation} for workloads 1 to 4, using SD-Policy DynAVGSD, to estimate a lower and upper bounds for this overhead.

\begin{figure}[htbp]
 \centering
 \includegraphics[width=1\columnwidth]{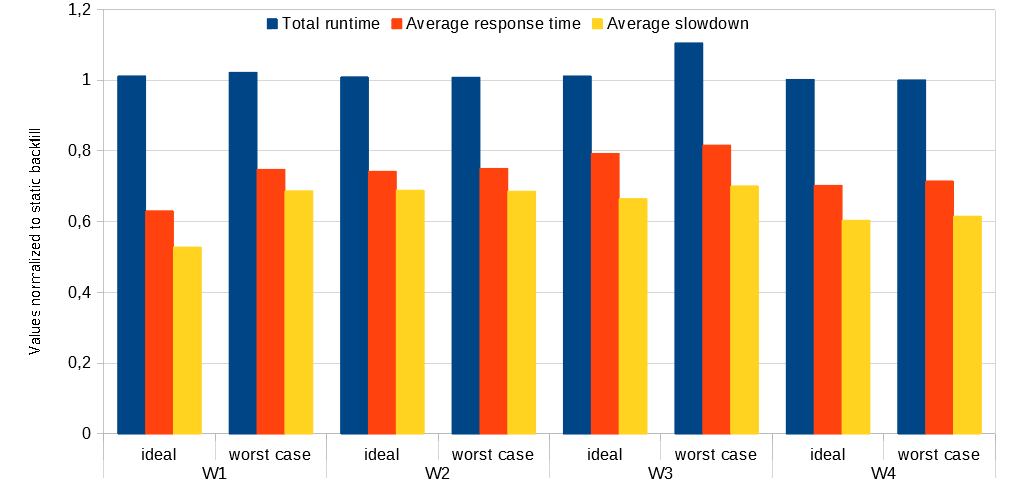}
 \caption{Makespan, average response time and slowdown for workloads 1-4, running SD-Policy with feedback, using ideal and worst case model. x axis represent the workloads, metrics are normalized to static backfill simulations.}
 \label{fig:idealvswc}
\vspace{-0.3cm}
\end{figure}

Figure~\ref{fig:idealvswc} shows the impact in makespan, average response time and average slowdown for the two models.
The worst case model increases response time, up to 11\% for workload 1 with respect to the ideal model, negligible for workload 3 and 4, less than 1.5\%. Average slowdown, similarly to average response time, increase by 16\% in Workload 1, only 3.5\% and 1\% in Workloads 3 and 4, while still outperforming static backfill. Makespan increases by 9\% in Workload 3, while less than 1\% in the other cases. Workload 2 is not affected by using the worst case model, as the jobs requested times are exact and it allows the SD-Policy to avoid creating unbalances.

Using a predictive model that lower the gap between requested time and real runtime would reduce the evaluated overhead to 0, as the time predictions would be more precise.

\subsection{Analysis of a real run}\label{realrun}
Having a proposed policy implemented with evaluations in a real system is a complex task, many publications skip this part, only basing the evaluation on simulation results. This is particularly true in the case of malleability, where all the layers of HPC software stack need to be aware of the directly connected layers.

On one hand, simulations are essentials, as they allow evaluating a scheduling algorithm in a big system with large workloads, but they do not have the same precision and complexity of a real system. It is difficult, for instance, evaluating the energy consumption of the SD-Policy and compare it with a real system, as well as modeling a real runtime model for different malleable jobs. We put effort in developing a working SLURM version, and we made its code available~\cite{sdpolicycode}, and we used it to evaluate Workload 5, based on the Cirne model using ANL arrival pattern, with 2000 jobs. Jobs ask a maximum of 16 nodes, 768 cores per job, on a system of 49 nodes, 2352 cores. We converted Cirne log to a real job submission list for SLURM by using a set of malleable applications.
We selected parameters to respect runtime and requested resources, and we generated, using a script, a list of submissions, respecting arrival time. Table~\ref{table:apps_real_run} shows the list of used applications, the type and the percentage of jobs running them.

We run the workload in MN4 supercomputer by spawning a SLURM instance inside a job of the SLURM installation in MN4, to have exclusive access to a subset of 50 nodes of the machine to run the workload. The number of nodes and the makespan of this evaluation are constrained by system queue limits, 50 nodes for 48 hours, so we adapted Cirne model to it. The controller uses the first node, while the other nodes are the computing nodes.

\begin{figure}[htbp]
 \centering
 \includegraphics[width=0.9\columnwidth]{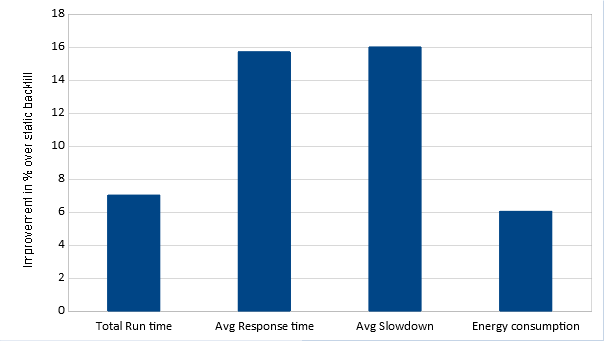}
 \caption{Improvement in percentage of the SD-Policy over the static backfill for makespan, average response time, average slowdown and energy consumption.}
 \label{fig:real_run}
 \vspace{-0.3cm}
\end{figure}

Results in Figure~\ref{fig:real_run} show that makespan decreases by 7\%, average response time and average slowdown by about 16\%, compared to the static backfill run. Results show less gain over the simulations because in this evaluation the considered number of nodes is lower. Shorter makespan is obtained by better utilization of nodes' resources, as 449 jobs out of 539 scheduled with malleability have a better runtime compared to the static execution, if we proportionate it to the number of used resources. This behavior is due two main reasons, already found in a previous research~\cite{D'Amico:2018:DEE:3229710.3229752}. The first reason is the better use of resources by memory bounded jobs, or jobs with memory bounded phases, i.e. during application's initialisation and finalization: in this situation, running computing-bounded jobs takes advantage of cores underutilized. The second reason is related to scalability problems of applications, that cannot perfectly scale to the high number of cores in the nodes: in this case, using malleability to partition available cores improves performance. As a consequence of better resources utilization, we save 6\% of energy over the static approach, very important result considering the increasing energy consumption of HPC systems.

\vspace{-0.3cm}
\section{Related work}\label{RW}
Malleability has been studied since many years ago in the context of scheduling, with the definition and study of Malleable Parallel Task Scheduling (MPTS) problem. The theoretical research shows its potential benefits~\cite{ludwig1994scheduling}~\cite{Turek:1994:SPT:181014.181331}~\cite{Mounie:1999:EAA:305619.305622}. These works are related to off-line schedulers, that select the number of resources that best improves the performance of the parallel task based on a model of its performance given at scheduling time, during the allotment. Once the task starts its execution, the number of its resources does not vary.

Today, the problem has been furtherer studied, Feitelson~\cite{10.1007/BFb0022284} classifies a malleable job as a job that can adapt to changes in the number of processors at run time. Following this definition, job scheduling simulations~\cite{1364755} showed the potential benefits of malleability on job's response time, by using fair process distribution and shrink and expand operation on jobs that expand the number of used processors.
Several studies have been done to reduce programming complexity when developing malleable applications. Utrera~\cite{10.1007/978-3-642-32820-6_20} uses folding techniques and a FCFS-malleable policy that uses co-scheduling and oversubscription to start MPI jobs when not enough resources are available.
Kale~\cite{1540460} implement malleable and evolving jobs on the top of Charm++, and defined a scheduling policy based on equipartition.
Prabhakaran~\cite{malsched:ipdps2015}, using Charm++ malleability, implemented shrink/expand operations in a production scheduler, together with a scheduling strategy based on equipartition and combining rigid, evolving and malleable jobs.
Martin~\cite{10.1007/978-3-642-40047-6_16} introduces FLEX-MPI library for dynamic reconfiguration of MPI applications based on checkpoint and restart, while Compr\'es~\cite{Compres:2016:IAE:2966884.2966917} implement MPI malleability with on-line data redistribution, plus shrink and expand operations in SLURM. In general, data redistribution in malleable MPI implies overhead of data movement and effort for developers, not in line with our research, being effortless and efficient.
Cera~\cite{Cera:2010:SMP:2018057.2018090} implements malleability based on dynamic CPUSETs using MPI and a production resource manager. This approach is similar to how we use malleability, but in our case, we do not oversubscribe MPI processes because we demonstrated it could degrade application's performance~\cite{djsb} and we integrate with shared memory programming models for better performance. While supporting MPI for multi-node applications, our approach uses DROM interface~\cite{D'Amico:2018:DEE:3229710.3229752}, that allows malleability in computational nodes by changing the number of threads OpenMP~\cite{OpenMPspec} or OmpSs~\cite{Duran2011OmpssAP} applications are using. This approach enables effortless, dynamic and zero overhead moldability and malleability in the number of cores application uses per node, and allows the scheduler to take decisions in real time with almost instantaneous adaptation from applications.

Many malleability approaches were studied in the context of the grid, taking advantage of application's feedback, like~\cite{6121261}~\cite{Vadhiyar:2005:SAG:1064323.1064329}~\cite{1195409}. Those approaches use application's performance models and run time performance measurements. Our algorithm differs, because it uses feedback from the scheduler itself, scheduling at a higher level of abstraction, based on system metrics, e.g., average system slowdown rather than application's feedback.

\section{Conclusions and future work}\label{future}
This paper presented a novel scheduling policy based on malleability and slowdown minimization, the SD-Policy. We described the algorithm and its implementation into a resource manager, enabling an integrating of the whole HPC software stack. We presented different parametric evaluations, together with two more complete cases of study, a big simulation and a real run into a supercomputer. We showed that SD-Policy can reduce the response time and the slowdown up to 50\% and 70\% for the CEA Curie log, while we saw an improvement of nodes utilization that brought 7\% makespan and 6\% energy reduction for a real run.

Future work will focus on three main points:
\begin{enumerate}
\item Integration of online performance metrics collection: by having information about applications we will be able to perform a better placement of malleable jobs, and dynamic adapting SharingFactor.
\item Improved scheduler: feedback mechanism, heuristic and runtime estimation, based on machine learning approaches.
\item Malleability at MPI level, unlocking new possibilities for the scheduler, that can shrink or expand jobs allocations in the number of nodes. 
\end{enumerate}

\begin{acks}
This work is partially supported by the Spanish Government through Programa Severo Ochoa (SEV-2015-0493), by the Spanish Ministry of Science and Technology through TIN2015-65316-P project, by the Generalitat de Catalunya (2017-SGR-1414) and from the European Union's Horizon 2020 under grant agreement No. 720270 (Human Brain Project SGA1) and No. 785907 (Human Brain Project SGA2).
\end{acks}
%
\bibliographystyle{ACM-Reference-Format}
\bibliography{sdpolicy.bib}

%



\end{document}